# Measurement of UV light emission of the nighttime Earth by Mini-EUSO for space-based UHECR observations

**K. Shinozaki*,[a] K. Bolmgren,[b] D. Barghini,[c,d] M. Battisti,[c,e] A. Belov,[f] M. Bertaina,[c,e] F. Bisconti,[e,c] G. Cambiè,[g,h] F. Capel,[b] M. Casolino,[g,i] F. Fenu,[c,d,e] A. Golzio,[c] P. Klimov,[f] V. Kungel,[j] L. Marcelli,[g] H. Miyamoto,[c,e] L. W. Piotrowski,[k] Z. Plebaniak,[c,e] M. Przybylak,[a] J. Szabelski,[a] N. Sakaki[g] and Y. Takizawa[g] on behalf of the JEM-EUSO Collaboration**

(a complete list of authors can be found at the end of the proceedings)

[a] *National Centre for Nuclear Research, Lodz, Poland*
[b] *KTH Royal Institute of Technology, Stockholm, Sweden*
[c] *Istituto Nazionale di Fisica Nucleare – Sezione di Torino, Turin, Italy*
[d] *Istituto Nazionale di Astrofisica – Osservatorio astrofisico di Torino, Turin, Italy*
[e] *Dipartimento di Fisica, Universitá di Torino, Turin, Italy*
[f] *Skobeltsyn Institute of Nuclear Physics, M.V. Lomonosov Moscow State University, Moscow, Russia*
[g] *Istituto Nazionale di Fisica Nucleare – Sezione di Roma Tor Vergata, Rome, Italy*
[h] *Universitá degli Studi di Roma Tor Vergata – Dipartimento di Fisica, Rome, Italy*
[i] *RIKEN, Wako, Japan*
[j] *Colorado School of Mines, Golden, USA*
[k] *University of Warsaw, Warsaw, Poland*

*E-mail:* kenji.shinozaki@ncbj.gov.pl

The JEM-EUSO (Joint Experiment Missions for Extreme Universe Space Observatory) program aims at the realization of the ultra-high energy cosmic ray (UHECR) observation using wide field of view fluorescence detectors in orbit. Ultra-violet (UV) light emission from the atmosphere such as airglow and anthropogenic light on the Earth's surface are the main background for the space-based UHECR observations. The Mini-EUSO mission has been operated on the International Space Station (ISS) since 2019 which is the first space-based experiment for the program. The Mini-EUSO instrument consists of a 25 cm refractive optics and the photo-detector module with the 2304-pixel array of the multi-anode photomultiplier tubes. On the nadir-looking window of the ISS, the instrument is capable of continuously monitoring a ~300 km × 300 km area. In the present work, we report the preliminary result of the measurement of the UV light in the nighttime Earth using the Mini-EUSO data downlinked to the ground. We mapped UV light distribution both locally and globally below the ISS obit. Simulations were also made to characterize the instrument response to diffuse background light. We discuss the impact of such light on space-based UHECR observations and the Mini-EUSO science objectives.

*Speaker

## 1. Introduction

The nature of the ultra-high energy cosmic rays (UHECRs) with energies $E_0$ above $\sim 10^{19}$ eV is an outstanding open question in astrophysics [1]. Their observation methods depend on the air shower phenomenon that is induced by the primary cosmic rays to result in a huge number of still relativistic secondary particles in the air shower. The fluorescence detectors (FDs) are employed by the state-of-the-art experiments to detect the UV fluorescence light originated by these secondary particles at night [2, 3]. Due to their extremely low fluxes, $\lesssim$1 km$^{-1}$ (1000 yr)$^{-1}$ for $E_0 > 10^{20}$ eV in particular, the measurements of UHECRs require a very large effective area of detection. Today, two leading air shower arrays are operated by the Pierre Auger Observatory [4], and the Telescope Array (TA) and its extension (TA×4) [5] with up to ~3000 km$^2$ effective areas, while despite of 100% duty cycle still larger statistics are needed particularly for the study of the UHECR origin.

The JEM-EUSO (Joint Experiment Missions for Extreme Universe Space Observatory) program aims at space-based UHECR observations [6]. By operating wide-FOV (field of view) FD(s) in orbit, UHECR-induced air shower events will be observed in as huge areas as $\sim 10^4$ – $10^5$ km$^2$. The program promotes the KLYPVE-EUSO mission [7] and jointly collaborates with the POEMMA (Probe of Extreme Multi-Messenger Astrophysics) project [8]. They will employ multi-square-meter scale telescope(s), i.e. FDs, with $\sim 45°$ FOV to be operated on the International Space Station (ISS) at ~400 km above sea level and on free-flier satellites at ~525 km, respectively.

To detect the air shower phenomena developing at speed of light in the lower atmosphere, spatially and temporarily high precision measurements are mandatory. Moreover, the signals of such events should be detected in the excess against background light. To fulfill the fundamental requirements, the collaboration has developed the photo-detector module (PDM) as a segment of the multi-channel ultra-fast detector placed on the focus [9]. Characteristics of such background were also studied in the experiments operating a PDM with a 1-m$^2$ Fresnel refractive optics on the ground [10] and on stratospheric balloons [11, 12] .

In the present work, we report the preliminary results of the similar background light study by the Mini-EUSO (Multiwavelength Imaging New Instrument for the Extreme Universe Space Observatory) mission, the first space-borne experiment for the collaboration [9]. The telescope was brought to the ISS in August 2019 and started operation in the following October. Same as the ground-based FDs, the airglow that emits at ~95 km height [13] acts as a persistent background component against UHECR observations. Orbiting Mini-EUSO allows the global study of background light including moonlight, and anthropogenic light, and clouds etc. In the following, we describe the instrument and data acquisition (DAQ) of Mini-EUSO, presents the UV light distribution in geographical location, i.e., maps, instrument response simulations, and discuss the role of such light for the future UHECR observations and the other scientific objectives.

## 2. Instrument and data acquisition

The ISS orbits with an inclination of $\sim 51.64°$. The mean height from the Earth's ellipsoid was ~420–430 km during the Mini-EUSO operation. In the same orbit, the height varies by $\sim \pm 13$ km. With a period of $\sim$ 93 min, the orbital speed is 7.6 km s$^{-1}$, while relative speed to the ground is reduced to be ~7.3 km s$^{-1}$ due to the Earth's rotation. The Mini-EUSO telescope is mounted on the nadir-looking UV transparent window of the *Zvezda* module at assigned time slots called 'sessions'. Up to July 2021, 41 sessions were completed. Each session happens once or twice a month and lasts ∼14 hours. A limited amount of the data were downlinked every session to assess correct functioning of DAQ. Now 43 hours' data are available on the ground including the entire data of ∼28 hours' DAQ in the first 15 sessions brought back by cosmonauts in May 2021.

The Mini-EUSO instrument consists of a refractive telescope and a PDM on the focus. The optics employs two flat Fresnel lenses made of UV-transparent PMMA (polymethyl methacrylate) with 250 mm diameter. The PDM is composed of 36 (=6×6) Hamamatsu R11265-103-M64 multi-anode photomultiplier tubes (MAPMTs). Each MAPMT has 64 (8 × 8) channels (pixels). The photocathode is glued with a BG3 band-pass filter to mainly accepts fluorescence light in the ∼330 – 400 nm wavelength. The non-zero transmittance extends in the ∼250 nm – 500 nm band. Near the center of the FOV, the angle of view of a pixel is ∼ 0.8°. It corresponds to a ∼ 5.9 km length on sea level seen from the typical ISS height. That of the PDM side is ∼ 44° corresponding to a ∼350 km length including the physical spaces between MAPMTs.

Readout electronics with SPACIROC-3 ASIC enable each channel for single photon counting every $\tau_{\mathrm{GTU}} = 2.5$ µs, referred to as a gate time unit (GTU). Mini-EUSO has three DAQ modes referred to as D1, D2, and D3 to observe various light emitting events that occur in the atmosphere with different time scales [9]. The D1 and D2 modes acquire the data by trigger algorithms [14] with 1 GTU and 128 GTUs time resolutions, respectively. The former is mainly used to test for the future observation of the UHECR-induced air showers. The latter is used to observe the slower events such as transient luminous events (TLEs). The D3 mode is used for the studies for even slower events such as meteor and strange quark matter search [9]. This mode continuously acquire the data by integrating counts over 16,384 (= 128 × 128) GTUs as a time frame over 40.96 ms. The sequential 128 frames are recorded as an event without interruption during the night in orbit. The D3 data are also used for even slower events such as meteors and strange quark matter search. Further details of the Mini-EUSO instrument can be referred to Ref. [9].

## 3. Analyses and preliminary results

In this section, we report the map and the absolute intensity of diffuse light from the analysis of the D3 data taking into account several corrections. Descriptions on key items are given followed by the preliminary results in the figures. Detailed interpretations are given in the next section.

The measured D3 count $n$ from photon counting were converted to the estimated number of photoelectrons (pe) $n_{\mathrm{pe}}$ per 1 GTU, Ref. [15] showed the output counts $n$ is to be

$$n = n_{\mathrm{pe}} \exp\left(-n_{\mathrm{pe}} \cdot \frac{\tau_0}{\tau_{\mathrm{GTU}}}\right), \qquad \text{where } \tau_0 = 5 \text{ ns double-pulse resolution.} \tag{1}$$

The discrepancy of $n$ from $n_{\mathrm{pe}}$ is ineligible for large counts, e.g., $n_{\mathrm{pe}} = 10$ (100) resulting in $n = 9.8$ (81). The maximum $n$ is ∼ 180 at $n_{\mathrm{pe}} \sim 500$. 'Flat-fielding' was applied to correct different efficiencies among pixels [14]. In the wide FOV, the optics shows a pincushion distortion. Based on the ray trace simulations, we estimated the correlation from the incoming photon direction to the expected focusing position on the PDM taken into account the spaces between MAPMTs. As the clock of the onboard controlling PC is not synchronized with the correct time, the time offset has been corrected every session by comparing the transit time of the known light sources like cities with the Night Earth map in the visible band provided by the DMSP satellite [16]. The attitude of the ISS whose Euler angles have a few degrees offset from the velocity vector and the zenith was separately corrected. Form the ISS position, these procedures correlate the geographical location in the FOV with the pixel of the PDM where originating photons are supposed to focus on. In orbit, it takes ∼48 s for the FOV to cross a particular position on the Earth's surface. Same locations are measured over ∼900 frames in the D3 mode excluding the insensitive area in the FOV.

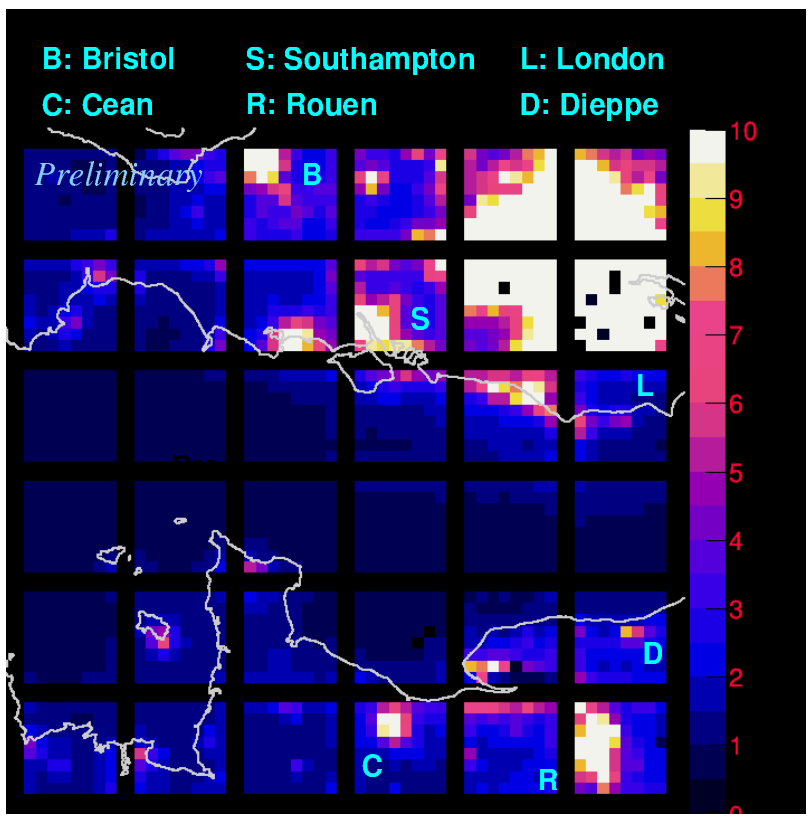

**Figure 1:** A D3 data image obtained above English Channel on the 48×48 pixels of the PDM. Labels are given to selected areas. Calculated coastlines are projected on the FOV. The color scale is in units of pe pixel$^{-1}$ GTU$^{-1}$.

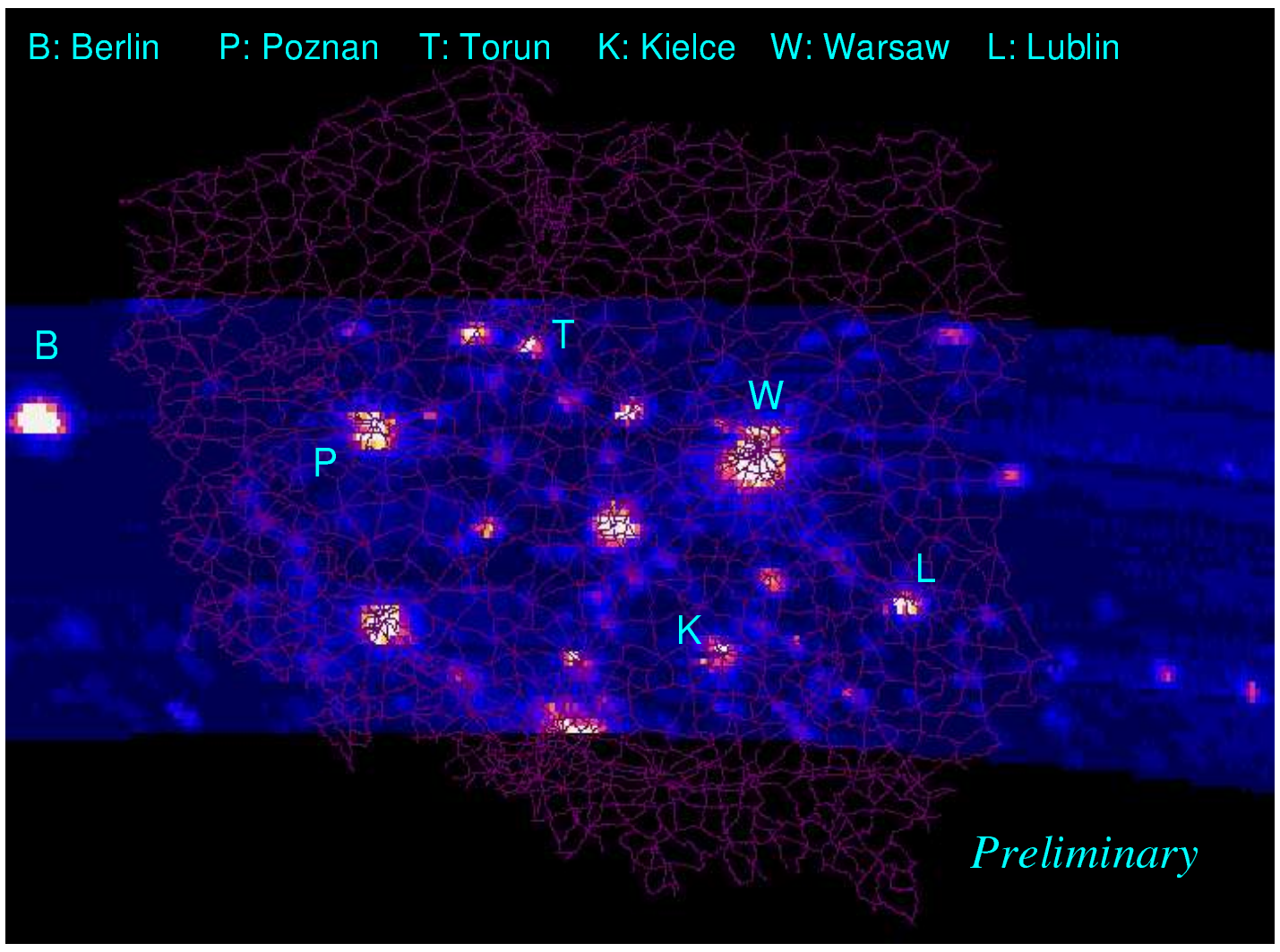

**Figure 2:** A combined peak-hold D3 image obtained in the same orbit ∼3 − 4 min after the DAQ of Figure 1 projected on 0.05° geographical grid points with the Polish road map of Poland. The same color scale is used.

Figure 1 displays an example of the D3 data image of a 40.96 ms frame obtained above English Channel projected on the 48×48 pixels of the PDM. The gaps represent physical spaces between MAPMTs. Labels are given for references. Calculated coastlines are projected on the FOV. The color scale is in units of pe GTU$^{-1}$.

Figure 2 shows an example of the combined image on 0.05° geographical grid points with the Polish road map using the peak-hold method. Data were obtained in the same orbit ∼3 − 4 min after the DAQ of Figure 1. The same color scale is used. Labels are given for references.

In addition, we investigated more global characteristics of the UV light measured for the selected sessions with proper flat-fielding. The ISS orbit covers the latitudes within ±51.64°. Except for the regions near these latitude extremes, it is unlikely to observe the same location on the Earth in the same session due to the Earth's rotation. On the other hand, it is frequent to observe the same location multiple times in different sessions. These data may be combined as long as the locating scheme works as demonstrated in Figures 1 and 2. In general, observations in different sessions suffer from different conditions such as the presence of Moon in the sky with various phases and zenith angles, presence of clouds in the FOV etc. In some cases the PDM may be partly not functional. To map globally and consistently, proper selection and correction of the data are needed. Additional information such as weather satellite images and meteorological models was also analyzed Ref. [17]. In the present work, we aim at characterizing and correcting the moonlight impact by first evaluating the correlation of the count rate with the Moon phase.

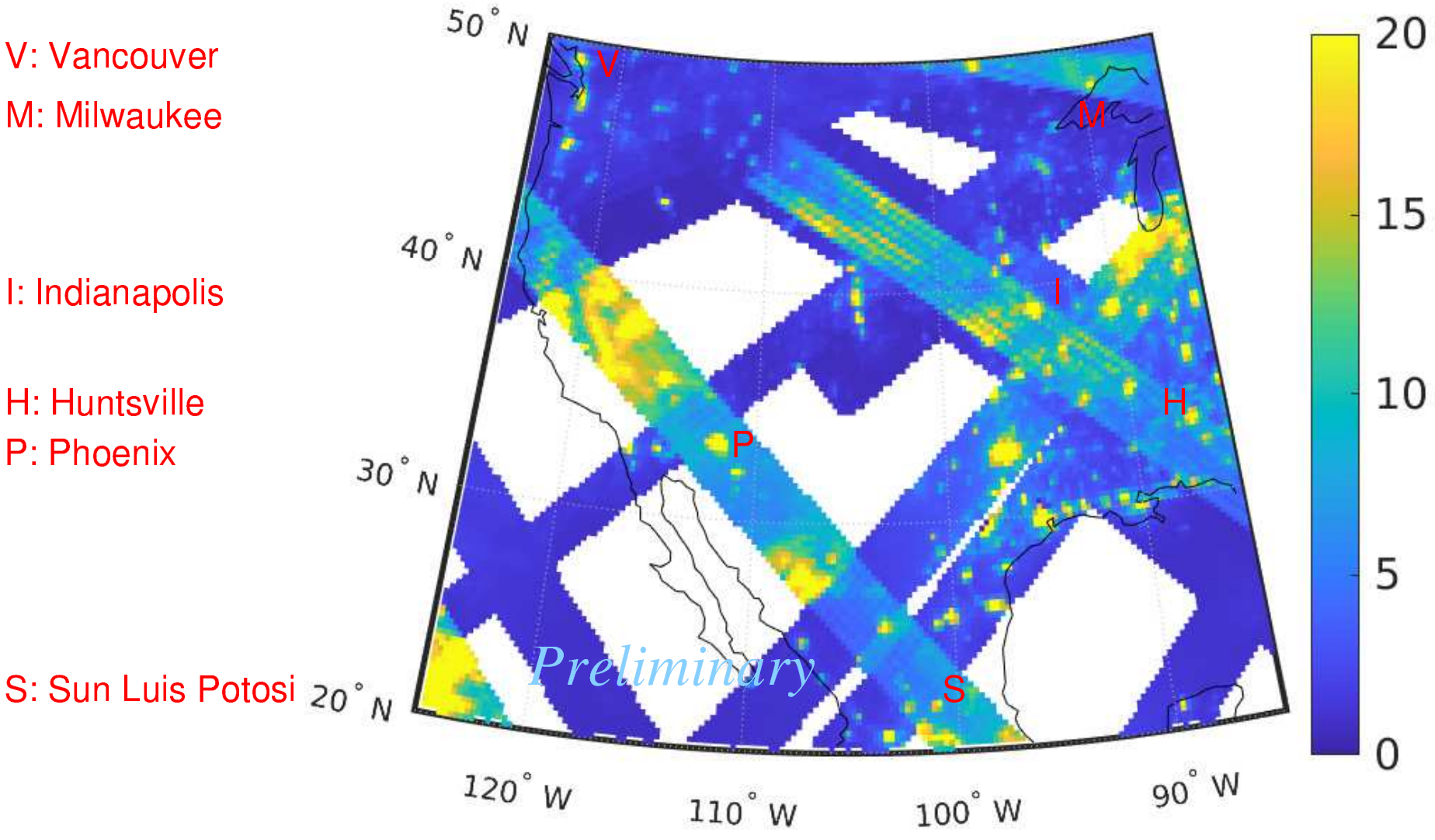

**Figure 3:** Combined image of the D3 data of multiple sessions on 0.25° geographical grid points over North America. The count rates in units of pe pixel$^{-1}$ GTU$^{-1}$ were corrected for moonlight impact and averaged.

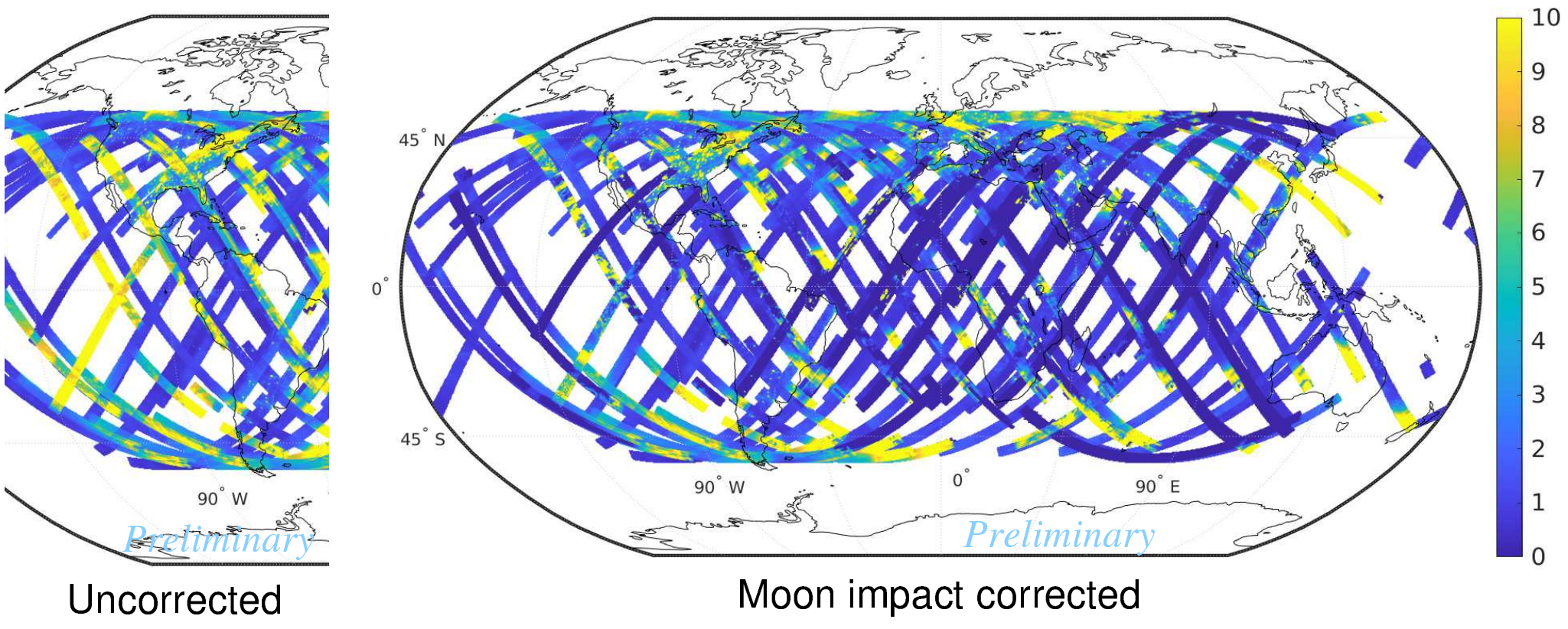

**Figure 4:** *Left*: uncorrected (part); and *Right*: moonlight-impact corrected count rate average on the global 1° geographical grid points.

Figure 3 shows an example of the combined image of the D3 data obtained above North America. The count rates in units of pe pixel$^{-1}$ GTU$^{-1}$ were corrected for moonlight impact using the above mentioned correlation and averaged on 0.25° geographical grid points.

Figure 4 compares uncorrected (*left*; partly for Americas) and moon impact corrected (*right*) average count rates on the global 1° geographical grid points.

In most cases, the measured count rates result from the unknown intensity $I_0$ of diffuse light with an unknown spectrum. In the present work, we deduced the consistent $I_0$ values for the measured count rates under the moonless conditions above the ocean where only diffuse background light is expected with no or little anthropogenic light in the FOV according to Ref. [11]. Under such conditions, the median of the count rates was $\tilde{n} \approx 2$ pe pixel$^{-1}$ GTU$^{-1}$ for the analyzed data.

The expected count rate $\bar{n}$ for a given $I_0$ value is given by

$$\bar{n} \approx I_0 \cdot \tau_{\mathrm{GTU}} \cdot \int_{\lambda} \left[ \dot{A}\,(\lambda) \cdot \frac{dF\,(\lambda)}{d\lambda} \right] d\lambda \tag{2}$$

where $dF\,(\lambda)/d\lambda$ is the fractional spectrum of the diffuse light within the band of interest. Using instrument response simulations, the 'acceptance' $\dot{A}_i$ of the $i$-th pixel to diffuse light is estimated by the ratio of the converted to photoelectrons $n_{\mathrm{pe},i}$ on that pixel to $N_{\mathrm{sim}}$ photons with a wavelength $\lambda$ uniformly injected on the lens area $S_0$ up to the large off-axis angle $\vartheta_{\max}$ beyond the FOV as follows:

$$\dot{A}_i\,(\lambda) = S_0 \cdot \left(\pi \cdot \sin^2 \vartheta_{\max}\right) \cdot \frac{n_{\mathrm{pe},i}}{N_{\mathrm{sim}}}. \tag{3}$$

The diffuse light in such quiet conditions is considered to be unknown mixtures of the light from natural origin. For $dF\,(\lambda)/d\lambda$ values, we assumed the modeled spectrum for airglow from Ref. [18] and starlight from Ref. [13]. The spectrum-weighted acceptance $\langle \dot{A} \rangle$ i.e., the convolution of $\dot{A}\,(\lambda)$ with $dF(\lambda)/d\lambda$, is a proportional constant from intensity to count rate.

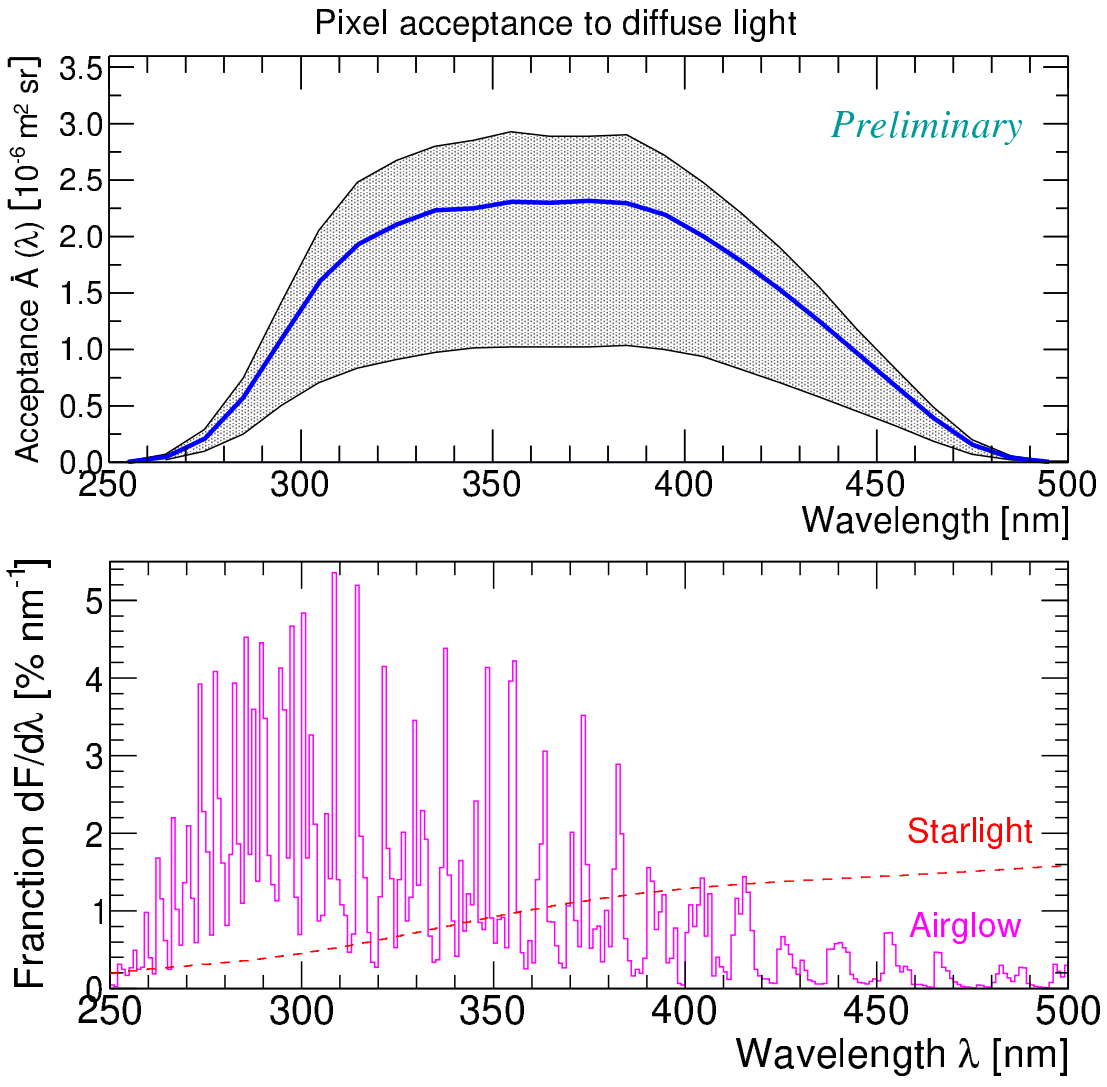

**Figure 5:** *Top*: Pixel acceptance to diffuse light as a function of wavelength. The solid curve and bounds of shaded indicates the average, maximum and minimum values in the PDM, respectively. *Bottom*: Assumed spectrum models for airglow [18] and starlight [13].

Figure 5 summarizes the instrument response simulation for diffuse light. The acceptance to diffuse light is as a function of wavelength. The solid curve and shaded region indicate the average, and maximum and minimum values among all pixels, respectively (*top*). The assumed input spectrum models for airglow [18] and starlight [13] (*bottom*). Convolving with the average $\dot{A}(\lambda)$ function, these models yield $\langle \dot{A} \rangle$ values to be $1.6 \times 10^{-6}$ m$^2$ sr and $1.3 \times 10^{-6}$ m$^2$ sr, respectively.

## 4. Discussion and summary

The JEM-EUSO program aims at the space-based UHECR observation and collaboration have studied the role of UV background light for example by EUSO-Balloon [11] that flew ~100 km distance at ~38 km height for one night. The present work by Mini-EUSO is the starting point for the same kind of analysis by the space-borne. Operating above the height of airglow emission, similar background light conditions to those of the future UHECR missions are expected allowing to investigate global and long-term characteristics.

Figures 1 to 3 show good performance for locating the geographical positions. Compared with remote areas, various sizes of populated areas are well recognized as in Figure 2. Clear contrast on the coastlines is also seen in Figures 1 and 3. This performance is one of the essential requirements not only for mapping but also for the air shower analysis to determine the geometry of the individual events and for evaluation of the exposure of UHECR observations from above. It also applies to the Mini-EUSO science objectivessuch as observations of TLEs and meteors, and search for strange quark matter [9]. As shown in Figures 3 and 4, characterization and correction of the moonlight impact on the count rates are underway. There were still areas with high count rates on the ocean, while the land part is generally moderated by this correction. In addition to the cloud impact [17], these effects need to be clarified to create the global map below the ISS orbit in the consistent conditions. As of July 2021, the majority of the recently acquired data since May 2020 is still on the ISS. Their retrieval is foreseen in a few months allowing to fill the unobserved areas in Figure 4 and more comprehensive studies about characteristics of such background light.

For any FDs for UHECR observation, small acceptance to diffuse light is highly desired to reduce the background noise by optimizing the optics and spectral response of the instrument, while keeping the high efficiency for the point-like sources. For the analysis of the physical events like air showers, it is more important to determine the count rates of the background than its absolute intensity. Reconstruction of the events requires the absolute calibration for the point-like sources with known fluorescence spectrum to evaluate the absolute signal flux from the detected excess signals with a high signal-to-noise ratio in terms of count rates. The absolute intensity of the background light is less relevant for such event analysis, however, it is still a valuable piece of information in particular to estimate the performance of the instrument before the mission. With the present knowledge and assumption on the instrument performance, the moonless count rate on the sea may be interpreted as an intensity $I_0 \sim 500 - 700$ m$^{-2}$ sr$^{-1}$ ns$^{-1}$ in the 250 – 500 nm band obtained for the models in the bottom panel of Figure 5. These models were simplified in comparison with Ref. [11] to evaluate the $I_0$ range. In accordance with progress of the Mini-EUSO mission, these value are subject to be updated.

It should be noted that, rather than measurement of the diffuse background light, the Mini-EUSO instrument is more optimal for detection of light signals from point-like sources. Figures 1 to 4 are good examples to demonstrate such performance. As seen in the top panel of Figure 5 the acceptance to diffuse light varies about by a factor of ~3 between the center and corners of the PDM due to optical vignetting. This method is intended for the diffuse light source with constant intensity as expected under the limited conditions. The measured counts are not always proportional to the diffuse intensity particularly for anthropogenic light whose spectrum is predominant in the visible band. Apart from cloud impact, different instruments, etc., the $I_0$ values from the present work are about twice values the corresponding to EUSO-Balloon [11]. Along with a variation of the airglow intensity, this larger intensity may be due to the airglow emission in the FOV.

In the present work, we reported the preliminary results on the measurement of the night Earth's UV light emission by Mini-EUSO. In the following months, more data will be available and the knowledge on the instrument performance will be upgraded by the planned calibration efforts [21]. In the text, we omitted reporting the temporal variation of the background light, which is a decisive factor for the observable threshold energy of air showers and observation time [19]. The present work, even focusing on the mapping, will contribute to the next balloon-borne mission, EUSO-SPB2 [20], and to estimate the performance of the future UHECR observation missions.

## Acknowledgments

This work was supported by State Space Corporation ROSCOSMOS, by the Italian Space Agency through the ASI INFN agreement n. 2020-26-HH.0 and contract n. 2016-1-U.0, by the French space agency CNES, National Science Centre in Poland grants 2017/27/B/ST9/02162 and 2020/37/B/ST9/01821. This research has been supported by the Interdisciplinary Scientific and Educational School of Moscow University "Fundamental and Applied Space Research". The article has been prepared based on research materials carried out in the space experiment "UV atmosphere".

## Full Authors List: JEM-EUSO

G. Abdellaoui[ah], S. Abe[fq], J.H. Adams Jr.[pd], D. Allard[cb], G. Alonso[md], L. Anchordoqui[pe], A. Anzalone[eh,ed], E. Arnone[ek,el], K. Asano[fe], R. Attallah[ac], H. Attoui[aa], M. Ave Pernas[mc], M. Bagheri[ph], J. Baláz[la], M. Bakiri[aa], D. Barghini[el,ek], S. Bartocci[ei,ej], M. Battisti[ek,el], J. Bayer[dd], B. Beldjilali[ah], T. Belenguer[mb], N. Belkhalfa[aa], R. Bellotti[ea,eb], A.A. Belov[kb], K. Benmessai[aa], M. Bertaina[ek,el], P.F. Bertone[pf], P.L. Biermann[db], F. Bisconti[el,ek], C. Blaksley[ft], N. Blanc[oa], S. Blin-Bondil[ca,cb], P. Bobik[la], M. Bogomilov[ba], K. Bolmgren[na], E. Bozzo[ob], S. Briz[pb], A. Bruno[eh,ed], K.S. Caballero[hd], F. Cafagna[ea], G. Cambié[ei,ej], D. Campana[ef], J-N. Capdevielle[cb], F. Capel[de], A. Caramete[ja], L. Caramete[ja], P. Carlson[na], R. Caruso[ec,ed], M. Casolino[ft,ei], C. Cassardo[ek,el], A. Castellina[ek,em], O. Catalano[eh,ed], A. Cellino[ek,em], K. Černý[bb], M. Chikawa[fc], G. Chiritoi[ja], M.J. Christl[pf], R. Colalillo[ef,eg], L. Conti[en,ei], G. Cotto[ek,el], H.J. Crawford[pa], R. Cremonini[el], A. Creusot[cb], A. de Castro Gónzalez[pb], C. de la Taille[ca], L. del Peral[mc], A. Diaz Damian[cc], R. Diesing[pb], P. Dinaucourt[ca], A. Djakonow[ia], T. Djemil[ac], A. Ebersoldt[db], T. Ebisuzaki[ft], J. Eser[pb], F. Fenu[ek,el], S. Fernández-González[ma], S. Ferrarese[ek,el], G. Filippatos[pc], W.I. Finch[pc] C. Fornaro[en,ei], M. Fouka[ab], A. Franceschi[ee], S. Franchini[md], C. Fuglesang[na], T. Fujii[fg], M. Fukushima[fe], P. Galeotti[ek,el], E. García-Ortega[ma], D. Gardiol[ek,em], G.K. Garipov[kb], E. Gascón[ma], E. Gazda[ph], J. Genci[lb], A. Golzio[ek,el], C. González Alvarado[mb], P. Gorodetzky[ft], A. Green[pc], F. Guarino[ef,eg], C. Guépin[pl], A. Guzmán[dd], Y. Hachisu[ft], A. Haungs[db], J. Hernández Carretero[mc], L. Hulett[pc], D. Ikeda[fe], N. Inoue[fn], S. Inoue[ft], F. Isgrò[ef,eg], Y. Itow[fk], T. Jammer[dc], S. Jeong[gb], E. Joven[me], E.G. Judd[pa], J. Jochum[dc], F. Kajino[ff], T. Kajino[fi], S. Kalli[af], I. Kaneko[ft], Y. Karadzhov[ba], M. Kasztelan[ia], K. Katahira[ft], K. Kawai[ft], Y. Kawasaki[ft], A. Kedadra[aa], H. Khales[aa], B.A. Khrenov[kb], Jeong-Sook Kim[ga], Soon-Wook Kim[ga], M. Kleifges[db], P.A. Klimov[kb], D. Kolev[ba], I. Kreykenbohm[da], J.F. Krizmanic[pf,pk], K. Królik[ia], V. Kungel[pc], Y. Kurihara[fs], A. Kusenko[fr,pe], E. Kuznetsov[pd], H. Lahmar[aa], F. Lakhdari[ag], J. Licandro[me], L. López Campano[ma], F. López Martínez[pb], S. Mackovjak[la], M. Mahdi[aa], D. Mandát[bc], M. Manfrin[ek,el], L. Marcelli[ei], J.L. Marcos[ma], W. Marszał[ia], Y. Martín[me], O. Martinez[hc], K. Mase[fa], R. Matev[ba], J.N. Matthews[pg], N. Mebarki[ad], G. Medina-Tanco[ha], A. Menshikov[db], A. Merino[ma], M. Mese[ef,eg], J. Meseguer[md], S.S. Meyer[pb], J. Mimouni[ad], H. Miyamoto[ek,el], Y. Mizumoto[fi], A. Monaco[ea,eb], J.A. Morales de los Ríos[mc], M. Mastafa[pd], S. Nagataki[ft], S. Naitamor[ab], T. Napolitano[ee], J. M. Nachtman[pi] A. Neronov[ob,cb], K. Nomoto[fr], T. Nonaka[fe], T. Ogawa[ft], S. Ogio[fl], H. Ohmori[ft], A.V. Olinto[pb], Y. Onel[pi] G. Osteria[ef], A.N. Otte[ph], A. Pagliaro[eh,ed], W. Painter[db], M.I. Panasyuk[kb], B. Panico[ef], E. Parizot[cb], I.H. Park[gb], B. Pastircak[la], T. Paul[pe], M. Pech[bb], I. Pérez-Grande[md], F. Perfetto[ef], T. Peter[oc], P. Picozza[ei,ej,ft], S. Pindado[md], L.W. Piotrowski[ib], S. Piraino[dd], Z. Plebaniak[ek,el,ia], A. Pollini[oa], E.M. Popescu[ja], R. Prevete[ef,eg], G. Prévôt[cb], H. Prieto[mc], M. Przybylak[ia], G. Puehlhofer[dd], M. Putis[la], P. Reardon[pd], M.H.. Reno[pi], M. Reyes[me], M. Ricci[ee], M.D. Rodríguez Frías[mc], O.F. Romero Matamala[ph], F. Ronga[ee], M.D. Sabau[mb], G. Saccá[ec,ed], G. Sáez Cano[mc], H. Sagawa[fe], Z. Sahnoune[ab], A. Saito[fg], N. Sakaki[ft], H. Salazar[hc], J.C. Sanchez Balanzar[ha], J.L. Sánchez[ma], A. Santangelo[dd], A. Sanz-Andrés[md], M. Sanz Palomino[mb], O.A. Saprykin[kc], F. Sarazin[pc], M. Sato[fo], A. Scagliola[ea,eb], T. Schanz[dd], H. Schieler[db], P. Schovánek[bc], V. Scotti[ef,eg], M. Serra[me], S.A. Sharakin[kb], H.M. Shimizu[fj], K. Shinozaki[ia], J.F. Soriano[pe], A. Sotgiu[ei,ej], I. Stan[ja], I. Strharský[la], N. Sugiyama[fj], D. Supanitsky[ha], M. Suzuki[fm], J. Szabelski[ia], N. Tajima[ft], T. Tajima[ft], Y. Takahashi[fo], M. Takeda[fe], Y. Takizawa[ft], M.C. Talai[ac], Y. Tameda[fp], C. Tenzer[dd], S.B. Thomas[pg], O. Tibolla[he], L.G. Tkachev[ka], T. Tomida[fh], N. Tone[ft], S. Toscano[ob], M. Traïche[aa], Y. Tsunesada[fl], K. Tsuno[ft], S. Turriziani[ft], Y. Uchihori[fb], O. Vaduvescu[me], J.F. Valdés-Galicia[ha], P. Vallania[ek,em], L. Valore[ef,eg], G. Vankova-Kirilova[ba], T. M. Venters[pj], C. Vigorito[ek,el], L. Villaseñor[hb], B. Vlcek[mc], P. von Ballmoos[cc], M. Vrabel[lb], S. Wada[ft], J. Watanabe[fi], J. Watts Jr.[pd], R. Weigand Muñoz[ma], A. Weindl[db], L. Wiencke[pc], M. Wille[da], J. Wilms[da], D. Winn[pm] T. Yamamoto[ff], J. Yang[gb], H. Yano[fm], I.V. Yashin[kb], D. Yonetoku[fd], S. Yoshida[fa], R. Young[pf], I.S. Zgura[ja], M.Yu. Zotov[kb], A. Zuccaro Marchi[ft]

[aa] Centre for Development of Advanced Technologies (CDTA), Algiers, Algeria

[ab] Dep. Astronomy, Centre Res. Astronomy, Astrophysics and Geophysics (CRAAG), Algiers, Algeria

[ac] LPR at Dept. of Physics, Faculty of Sciences, University Badji Mokhtar, Annaba, Algeria

[ad] Lab. of Math. and Sub-Atomic Phys. (LPMPS), Univ. Constantine I, Constantine, Algeria
[af] Department of Physics, Faculty of Sciences, University of M'sila, M'sila, Algeria
[ag] Research Unit on Optics and Photonics, UROP-CDTA, Sétif, Algeria
[ah] Telecom Lab., Faculty of Technology, University Abou Bekr Belkaid, Tlemcen, Algeria
[ba] St. Kliment Ohridski University of Sofia, Bulgaria
[bb] Joint Laboratory of Optics, Faculty of Science, Palacký University, Olomouc, Czech Republic
[bc] Institute of Physics of the Czech Academy of Sciences, Prague, Czech Republic
[ca] Omega, Ecole Polytechnique, CNRS/IN2P3, Palaiseau, France
[cb] Université de Paris, CNRS, AstroParticule et Cosmologie, F-75013 Paris, France
[cc] IRAP, Université de Toulouse, CNRS, Toulouse, France
[da] ECAP, University of Erlangen-Nuremberg, Germany
[db] Karlsruhe Institute of Technology (KIT), Germany
[dc] Experimental Physics Institute, Kepler Center, University of Tübingen, Germany
[dd] Institute for Astronomy and Astrophysics, Kepler Center, University of Tübingen, Germany
[de] Technical University of Munich, Munich, Germany
[ea] Istituto Nazionale di Fisica Nucleare - Sezione di Bari, Italy
[eb] Universita' degli Studi di Bari Aldo Moro and INFN - Sezione di Bari, Italy
[ec] Dipartimento di Fisica e Astronomia "Ettore Majorana", Universita' di Catania, Italy
[ed] Istituto Nazionale di Fisica Nucleare - Sezione di Catania, Italy
[ee] Istituto Nazionale di Fisica Nucleare - Laboratori Nazionali di Frascati, Italy
[ef] Istituto Nazionale di Fisica Nucleare - Sezione di Napoli, Italy
[eg] Universita' di Napoli Federico II - Dipartimento di Fisica "Ettore Pancini", Italy
[eh] INAF - Istituto di Astrofisica Spaziale e Fisica Cosmica di Palermo, Italy
[ei] Istituto Nazionale di Fisica Nucleare - Sezione di Roma Tor Vergata, Italy
[ej] Universita' di Roma Tor Vergata - Dipartimento di Fisica, Roma, Italy
[ek] Istituto Nazionale di Fisica Nucleare - Sezione di Torino, Italy
[el] Dipartimento di Fisica, Universita' di Torino, Italy
[em] Osservatorio Astrofisico di Torino, Istituto Nazionale di Astrofisica, Italy
[en] Uninettuno University, Rome, Italy
[fa] Chiba University, Chiba, Japan
[fb] National Institutes for Quantum and Radiological Science and Technology (QST), Chiba, Japan
[fc] Kindai University, Higashi-Osaka, Japan
[fd] Kanazawa University, Kanazawa, Japan
[fe] Institute for Cosmic Ray Research, University of Tokyo, Kashiwa, Japan
[ff] Konan University, Kobe, Japan
[fg] Kyoto University, Kyoto, Japan
[fh] Shinshu University, Nagano, Japan
[fi] National Astronomical Observatory, Mitaka, Japan
[fj] Nagoya University, Nagoya, Japan
[fk] Institute for Space-Earth Environmental Research, Nagoya University, Nagoya, Japan
[fl] Graduate School of Science, Osaka City University, Japan
[fm] Institute of Space and Astronautical Science/JAXA, Sagamihara, Japan
[fn] Saitama University, Saitama, Japan

[fo] Hokkaido University, Sapporo, Japan
[fp] Osaka Electro-Communication University, Neyagawa, Japan
[fq] Nihon University Chiyoda, Tokyo, Japan
[fr] University of Tokyo, Tokyo, Japan
[fs] High Energy Accelerator Research Organization (KEK), Tsukuba, Japan
[ft] RIKEN, Wako, Japan
[ga] Korea Astronomy and Space Science Institute (KASI), Daejeon, Republic of Korea
[gb] Sungkyunkwan University, Seoul, Republic of Korea
[ha] Universidad Nacional Autónoma de México (UNAM), Mexico
[hb] Universidad Michoacana de San Nicolas de Hidalgo (UMSNH), Morelia, Mexico
[hc] Benemérita Universidad Autónoma de Puebla (BUAP), Mexico
[hd] Universidad Autónoma de Chiapas (UNACH), Chiapas, Mexico
[he] Centro Mesoamericano de Física Teórica (MCTP), Mexico
[ia] National Centre for Nuclear Research, Lodz, Poland
[ib] Faculty of Physics, University of Warsaw, Poland
[ja] Institute of Space Science ISS, Magurele, Romania
[ka] Joint Institute for Nuclear Research, Dubna, Russia
[kb] Skobeltsyn Institute of Nuclear Physics, Lomonosov Moscow State University, Russia
[kc] Space Regatta Consortium, Korolev, Russia
[la] Institute of Experimental Physics, Kosice, Slovakia
[lb] Technical University Kosice (TUKE), Kosice, Slovakia
[ma] Universidad de León (ULE), León, Spain
[mb] Instituto Nacional de Técnica Aeroespacial (INTA), Madrid, Spain
[mc] Universidad de Alcalá (UAH), Madrid, Spain
[md] Universidad Politécnia de madrid (UPM), Madrid, Spain
[me] Instituto de Astrofísica de Canarias (IAC), Tenerife, Spain
[na] KTH Royal Institute of Technology, Stockholm, Sweden
[oa] Swiss Center for Electronics and Microtechnology (CSEM), Neuchâtel, Switzerland
[ob] ISDC Data Centre for Astrophysics, Versoix, Switzerland
[oc] Institute for Atmospheric and Climate Science, ETH Zürich, Switzerland
[pa] Space Science Laboratory, University of California, Berkeley, CA, USA
[pb] University of Chicago, IL, USA
[pc] Colorado School of Mines, Golden, CO, USA
[pd] University of Alabama in Huntsville, Huntsville, AL; USA
[pe] Lehman College, City University of New York (CUNY), NY, USA
[pf] NASA Marshall Space Flight Center, Huntsville, AL, USA
[pg] University of Utah, Salt Lake City, UT, USA
[ph] Georgia Institute of Technology, USA
[pi] University of Iowa, Iowa City, IA, USA
[pj] NASA Goddard Space Flight Center, Greenbelt, MD, USA
[pk] Center for Space Science & Technology, University of Maryland, Baltimore County, Baltimore, MD, USA
[pl] Department of Astronomy, University of Maryland, College Park, MD, USA
[pm] Fairfield University, Fairfield, CT, USA